\documentclass[aps,prd,twocolumn,showpacs,superscriptaddress,groupedaddress,nofootinbib,floatfix]{revtex4}

\usepackage{graphicx}% Include figure files
\usepackage{dcolumn}% Align table columns on decimal point
\usepackage{bm}% bold math
\usepackage{graphicx}	% Including figure files
\usepackage{amsmath}	% Advanced maths commands
\usepackage{amssymb}	% Extra maths symbols

\usepackage{epsfig,amsmath,natbib}
\usepackage{color,subfigure}
\usepackage{xcolor}

\usepackage{mathrsfs,amssymb,amstext}
\usepackage{ulem}%%%%%??????
\usepackage{url}
\usepackage{bm}
\usepackage{adjustbox}
\usepackage{threeparttable}

\newcommand{\Mpc}{\mathrm{~km~s^{-1}~Mpc^{-1}}}
\newcommand{\kms}{\mathrm{~km~s^{-1}}}

\begin{document}

%\preprint{APS/123-QED}

\title{Hubble constant from the cluster-lensed quasar system SDSS J1004+4112: investigation of the lens model dependence}% Force line breaks with \\

%\affiliation{%

%%%%%%%

\author{Yuting Liu\footnote{yutingl@mail.bnu.edu.cn}}
\affiliation{Institute for Frontiers in Astronomy and Astrophysics, Beijing Normal University, Beijing 102206, China}
\affiliation{Department of Astronomy, Beijing Normal University, Beijing 100875, China}
\affiliation{Department of Physics, University of Tokyo, Tokyo 113-0033, Japan}

\author{Masamune Oguri\footnote{masamune.oguri@chiba-u.jp}} 
\affiliation{Center for Frontier Science, Chiba University,  Chiba 263-8522, Japan}
\affiliation{Department of Physics, Graduate School of Science, Chiba University, Chiba 263-8522, Japan}
\affiliation{Kavli Institute for the Physics and Mathematics of the Universe, University of Tokyo, Kashiwa, Chiba 277-8583, Japan}

\author{Shuo Cao\footnote{caoshuo@bnu.edu.cn}} 
\affiliation{Institute for Frontiers in Astronomy and Astrophysics, Beijing Normal University, Beijing 102206, China}
\affiliation{Department of Astronomy, Beijing Normal University, Beijing 100875, China}

\date{\today}% It is always \today, today,

\begin{abstract}
As a fundamental parameter for modern cosmology, the Hubble constant $H_0$ is experiencing a serious crisis. In this paper, we explore an independent approach to measure $H_0$ based on the time-delay cosmography with strong gravitational lensing of a quasar by a galaxy cluster. Specifically we focus on the strong lensing system SDSS J1004+4112 with the maximum image separation of 14.62$''$, the first system of a quasar lensed by a galaxy cluster with five multiple images. Incorporating the latest time-delay measurements, we investigate the lens model dependence from the combination of 16 different lens mass models. We find that the lens model dependence is indeed large, with the combined measurement of the Hubble constant of $H_0=67.5^{+14.5}_{-8.9}\Mpc$ that is obtained by summing posteriors of the Hubble constant from the 16 models with equal weighting.  Interestingly, our results show that the value of Hubble constant decreases as the complexity of the perturbation around the lens increases, although weighting based on positional errors of quasar images does not significantly improve the $H_0$ constraint. We find that the 16 different mass models predict largely different shapes of the lensed quasar host galaxy as well as other lensed galaxies behind the cluster.  By selecting two mass models that best reproduces those shapes, the constraint on the Hubble constant is significantly tightened to  $H_0=59.1^{+3.6}_{-3.5}\Mpc$. While we caution that our analysis still does not fully explore all the possible mass model uncertainty, our results highlight the importance of including as many constraints as possible such as extended shapes of lensed galaxies for obtaining tight constraints on the Hubble constant from cluster-lensed quasar lens systems.

%One thing to note that this value should be treated with caution because we did not fully explore all the mass model uncertainty.
%Therefore, fine lens modeling will help us better understand the Hubble constant and improve the Hubble tension.

\end{abstract}

\keywords{cosmological parameters, gravitational lensing, quasars, galaxy clusters}
\maketitle

%\tableofcontents

\section{Introduction}\label{sec:intro}

The Hubble constant $H_0$, which quantifies the expansion rate of the Universe today, is one of the most important cosmological parameters that is also related to the age and the total energy density of the Universe \cite{2013PhR...530...87W,2001ApJ...553...47F}. 
%Therefore, a precise measurement of the Hubble constant is critical to the study of modern cosmology. 
Although $H_0$ has been measured with high precision, the determination of its accurate value is extremely challenging, followed with an increasing tension.
In particular, the discrepancy of measured $H_0$ values between different measurements is becoming more pronounced. The cosmic microwave background radiation of the early Universe from the \textit{Planck} satellite provides a tight constraint on the
Hubble constant as $H_{0}=67.4\pm0.5\Mpc$, which is based on the so-called standard cosmology, $\Lambda$-dominated cold dark matter ($\Lambda$CDM) model \cite{2020A&A...641A...6P}. Such concordance $\Lambda$CDM model has withstood most popular
observational evidences \cite{1998AJ....116.1009R,1999ApJ...517..565P,2015ApJ...806..185C,2018PhRvD..98d3528T,2018ApJ...859..101S,2020PhRvD.102h3504B,2021JCAP...01..031H,2021A&A...645A.104A,2020Natur.587..210M}. An independent measurement of $H_0$ has been obtained by type Ia supernovae (SNe Ia) calibrated via the distance ladder in the local Universe, with the value of $H_0=73.04\pm 1.04 \Mpc$ released by SH0ES (SNe, H0, for the Equation of State of dark energy) collaboration \cite{2022ApJ...934L...7R}. It is obvious that the $H_0$ constraints derived by these two methods are inconsistent. Such significant 5$\sigma$ tension  has sparked debates about the validity of the $\Lambda$CDM model.

As a consequence, any other methodology to measure the Hubble constant is very necessary to clarify the origin of the discrepancy. The tip of the red giant branch (TRGB) as a local standard candle provides another way to determine the Hubble constant. Ref.~\cite{2019ApJ...882...34F} presented $H_0=69.8\pm0.8 \Mpc$ by measuring TRGB in nine SNe Ia hosts and calibrating TRGB in the Large Magellanic Cloud. Several recent calibrations of the TRGB method were further combined, which are internally self-consistent at the 1\% level \cite{2021ApJ...919...16F}. Making geometric distance measurements to megamaser galaxies in the Hubble flow is
another particular approach to constrain the Hubble constant
\cite{2008ApJ...678...96B,2013ApJ...767..154R,2013ApJ...767..155K,2015ApJ...800...26K,2017ApJ...834...52G}. Recently, Ref.~\cite{2020ApJ...891L...1P} employed such approach to derive the value of $H_0=73.9\pm3.0 \Mpc$. In addition, gravitational waves from compact binary mergers provide a useful means of directly measuring luminosity distances, and therefore provide an independent approach to constrain $H_0$, which is sometimes referred to as the standard siren technique. For instance, 
from the neutron star merger event GW170817 with its electromagnetic counterparts AT2017gfo and GRB170817A,  Ref.~\cite{2020Sci...370.1450D} obtained the Hubble constant value of $H_0=66.2^{+4.4}_{-4.2}\Mpc$. The LIGO-Virgo-KAGRA Scientific Collaboration obtained $H_0=68^{+12}_{-8} \Mpc$ with 47 gravitational-wave sources from the Third LIGO-Virgo-KAGRA Gravitational-Wave Transient Catalog (GWTC3) \cite{2021arXiv211103604T}. 

It is extremely important to further derive measurements of $H_0$ using independent techniques such as strong gravitational lensing. Ref.~\cite{1964MNRAS.128..307R} proposed that observations of time delays between multiple images of gravitationally lensed supernovae could be used to determinate $H_0$ independently of the distance ladder. Very recently, this method finally applied to the gravitationally lensed supernova SN Refsdal to obtain the constraint on the Hubble constant of $64.8^{+4.4}_{-4.3} \Mpc$ \cite{2023Sci...380.1322K}. 

Given the rarity of lensed supernovae, lensed quasars have so far been mostly used to constrain the Hubble constant.  For instance, the H0LiCOW ($H_0$ Lenses in COSMOGRAIL's Wellspring) collaboration presented the estimation of the Hubble constant from the analysis of time delays for six lensed quasars as $H_0=73.3^{+1.7}_{-1.8} \Mpc$ \cite{2020MNRAS.498.1420W}. While the H0LiCOW result was obtained by assuming rather simplistic radial mass distributions for lensing galaxies,  the TDCOSMO (Time-Delay COSMOgraphy) team addressed the effect of lens model degeneracies by considering more flexible radial mass distributions and breaking the degeneracies by stellar kinematics, yielding a somewhat weaker constraint of $H_0=74.5^{+5.6}_{-6.1} \Mpc$ \cite{2020A&A...643A.165B}. In the same paper, a slightly different value of $H_0=67.4^{+4.1}_{-3.2} \Mpc$ was also obtained by adding a prior from Sloan ACS survey galaxy-galaxy strong lens systems. Recently, Ref.~\cite{2023A&A...673A...9S} obtained  $H_0=77.1^{+7.3}_{-7.1} \Mpc$ from the lens RXJ1131-1231 using spatially resolved stellar kinematics of the lens galaxy. Clearly a challenge lies in how to break degeneracies between different mass models for obtaining tight constraints on the Hubble constant.

We note that galaxy-scale lensed quasars are used in the H0LiCOW and TDCOSMO analysis. Since lens modeling of galaxy- and cluster-scale lenses involve quite different systematic errors, it is worth exploring the possibility of obtaining the complementary constraint on the Hubble constant from cluster-lensed quasar systems, which is a relatively new area of research partly because of the small sample of such strong lens systems. One notable difference between galaxy- and cluster-scale lensed quasars is that constraints on the lens mass model from additional multiple images of galaxies behind the lensing cluster are commonly available for the latter \cite{2005ApJ...629L..73S,2017ApJ...835....5S,2022ApJ...926...86A,2022A&A...668A.142A}. 
While it is known that cluster strong lens mass modeling exhibits relatively large positional offsets between observed and model-predicted multiple images, a careful analysis with mock strong lens data from ray-tracing simulations \cite{2017MNRAS.472.3177M} has demonstrated that it is possible to recover input Hubble constant accurately even with such large positional offsets \cite{2023Sci...380.1322K}.
Ref.~\cite{2021MNRAS.501..784D} used the cluster-scale lensed quasar SDSS J1004+4112 \cite{2003Natur.426..810I,2004ApJ...605...78O,2005PASJ...57L...7I,2008ApJ...676..761F} together with 7 galaxy-scale lensed quasars to obtain $H_0=71.8^{+3.9}_{-3.3} \Mpc$. Ref.~\cite{2023arXiv230111240N} focused more specifically on cluster-scale lensed quasars by studying three cluster-lensed quasars SDSS J1004+4112, SDSS J1029+2623 \cite{2006ApJ...653L..97I,2008ApJ...676L...1O,2013ApJ...764..186F}, and SDSS J2222+2745 \cite{2013ApJ...773..146D,2015ApJ...813...67D}, to produce a combined measurement of Hubble constant as $H_0=71.5\pm6.1\Mpc$. However, in this analysis only one specific mass model is considered for each lens system, and as a result the degeneracy among different mass models has not been fully explored.
Very recently, Ref.~\cite{2023arXiv230914776M}
presented a 
new constraint on the Hubble constant from the 
SDSS J1004 + 4112, $H_0=74^{+9}_{-13}\Mpc$, using a free-form lens model.

In this paper, we focus on the first cluster-lensed quasar system SDSS J1004+4112 \cite{2003Natur.426..810I,2004ApJ...605...78O,2005PASJ...57L...7I} to derive an independent $H_0$ measurement, paying a careful attention to the dependence of the $H_0$ constraint on the assumed lens mass model. Specially, the latest time-delay measurements between different images \cite{2022ApJ...937...34M} and the combination of 16 different lens mass models are considered in order to address the lens mass dependence on derived values of $H_0$. Our work is complementary to Ref.~\cite{2022ApJ...937...35F} in which the latest time-delay measurements are used to place tight constraints on the lens mass distribution. We also discuss possibilities to break the lens model degeneracy by adding other observables to obtain better constraints on $H_0$. Our work will be useful for future cosmological application of cluster-lensed quasar systems that are actively being searched and discovered recently \cite{2018MNRAS.481L.136S,2019MNRAS.489.4741S,2023ApJ...946...63M,2023arXiv230514317N}. 

This paper is arranged as follows. In Sec.~\ref{sec:data}, we briefly describe the cluster-lensed quasar system SDSS J1004+4112. We introduce the methodology of lens modeling in Sec.~\ref{sec:model}.
Results of our analysis are presented in Sec.~\ref{sec:result}. The possible improvement of the Hubble constant with additional constraints is discussed in Sec.~\ref{sec:discussion}. We summarize the main conclusion in Sec.~\ref{sec:summary}.

\begin{figure}
\begin{center}
\includegraphics[width=0.95\linewidth]{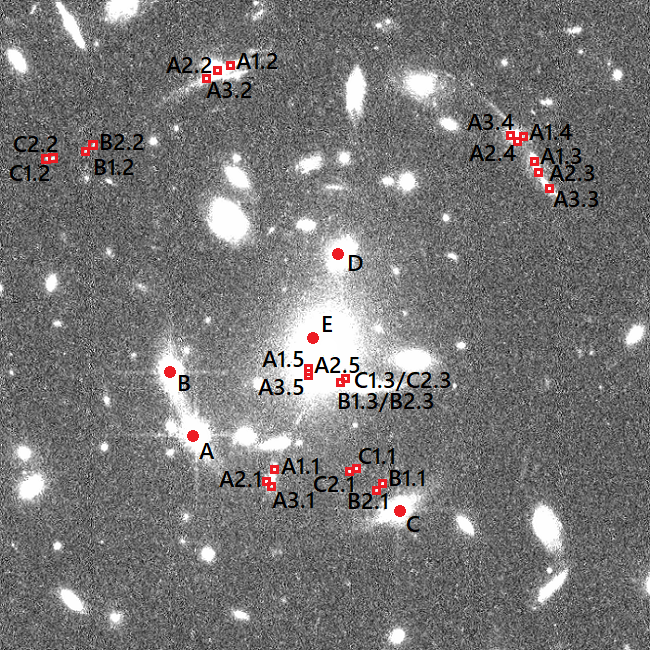}
\end{center}
\caption{The HST/ACS F814W image of the cluster-lensed quasar system SDSS J1004+4112. Red dots labeled with A-E are five multiple images of a lensed quasar.
Red squares are
images of three lensed background galaxies at redshifts 3.33, 2.74, and 3.28, respectively.}
\label{Fig1}
\end{figure}

\begin{table}
  \caption{Observational constraints from multiple images of lensed quasar and galaxies.
  Source redshift, $x$- and $y$-positions with respect to the quasar image A, the magnitude difference, and the relative time delay are listed in columns 2--5, respectively.
  \label{table:1}}
  \vspace{3mm}
   %\begin{center}
  \setlength{\tabcolsep}{1.2mm}{
    \begin{tabular}{cccccc}
     \hline\hline
      Name & $z_{\rm s}$ &$\Delta x$ [$''$] & $\Delta y $ [$''$] & $\Delta m $ &
      $\Delta t$ [days]\\
     \hline
A &  1.734& 0.000   & 0.000  &  $\equiv  0$ & $825.99\pm2.10$  \\
B &       &$-1.317$ &3.532  &  $0.35\pm 0.3$ & $781.92\pm2.20$ \\
C &       &11.039   &$-4.492$&  $0.87\pm 0.3$ & $\equiv  0$\\
D &       &8.399    & 9.707  &  $1.50\pm 0.3$ &  $2456.99\pm5.55$     \\
E &       &7.197    & 4.603  &  $6.30\pm 0.8$ &  $\cdots$     \\
      \hline
A1.1 & 3.33 & 3.93  & $-2.78$ \\
A1.2 &      & 1.33  & 19.37  \\
A1.3 &      &19.23  & 14.67  \\
A1.4 &      &18.83  & 15.87  \\
A1.5 &      & 6.83  &  3.22  \\
   \hline
A2.1 & 3.33 &4.13   &$-2.68$  \\
A2.2 &      &   1.93&  19.87  \\
A2.3 &      &  19.43&  14.02  \\
A2.4 &      &  18.33&  15.72  \\
A2.5 &      &   6.83&   3.12  \\
   \hline
A3.1 & 3.33 & 4.33  & $-1.98$ \\
A3.2 &      & 2.73  & 20.37  \\
A3.3 &      &19.95  & 13.04  \\
A3.4 &      &18.03  & 15.87  \\
A3.5 &      & 6.83  &  3.02  \\
   \hline
B1.1 & 2.74 & 8.88  & $-2.16$ \\
B1.2 &      &$-5.45$& 15.84 \\
B1.3 &      & 8.33  &  2.57 \\
   \hline
B2.1 & 2.74 & 8.45  & $-2.26$\\
B2.2 &      &$-5.07$& 16.04 \\
B2.3 &      &  8.33 &  2.57 \\
   \hline
C1.1 & 3.28 & 10.25 &$-3.06$ \\
C1.2 &      &$-7.55$& 15.39 \\
C1.3 &      &  8.49 &  2.72 \\
   \hline
C2.1 & 3.28 & 9.95  & $-3.36$\\
C2.2 &      &$-7.30$& 15.44 \\
C2.3 &      &  8.49 &  2.72 \\
  \hline\hline
%\multicolumn{6}{@{}l@{}}{\hbox to 0pt{\parbox{85mm}\hss}}
\end{tabular}}
  %\end{center}
\end{table}

\section{Observations of SDSS J1004+4112} \label{sec:data}

\begin{table}
  \caption{Assumed positional errors of quasar images $\sigma_{\rm q}$ and galaxy images $\sigma_{\rm g}$ for the 16 different lens mass models.} 
  \label{table:2}
  \vspace{3mm}
  %\begin{center}
  \begin{threeparttable}
  \setlength{\tabcolsep}{0.4mm}{
    \begin{tabular}{cccc}
     \hline\hline
 Label &   Model \tnote{*}    &  $\sigma_{\rm q}$[$''$]   & $\sigma_{\rm g}$[$''$]   \\\\
     \hline
m111 &{\tt anfw}+{\tt gals}+{\tt pert}  &0.12 & 1.2 \\
m112 &{\tt anfw}+{\tt gals}+{\tt pert}+{\tt mpole}(m=3)  &0.07 & 0.7 \\
m113 &{\tt anfw}+{\tt gals}+{\tt pert}+{\tt mpole}(m=3,4)   &0.05 & 0.5 \\
m114 &{\tt anfw}+{\tt gals}+{\tt pert}+{\tt mpole}(m=3,4,5) &0.04 & 0.4 \\
     \hline
m121 &{\tt anfw}+{\tt ahern}+{\tt gals}+{\tt pert}  &0.07 & 0.7 \\
m122 &{\tt anfw}+{\tt ahern}+{\tt gals}+{\tt pert}+{\tt mpole}(m=3)  &0.05 & 0.5  \\
m123 &{\tt anfw}+{\tt ahern}+{\tt gals}+{\tt pert}+{\tt mpole}(m=3,4)  &0.04 & 0.4 \\
m124 &{\tt anfw}+{\tt ahern}+{\tt gals}+{\tt pert}+{\tt mpole}(m=3,4,5) &0.03 & 0.3 \\
     \hline
m211 &{\tt jaffe}+{\tt gals}+{\tt pert}   &0.12 & 1.2 \\
m212 &{\tt jaffe}+{\tt gals}+{\tt pert}+{\tt mpole}(m=3)  &0.09 & 0.9 \\
m213 &{\tt jaffe}+{\tt gals}+{\tt pert}+{\tt mpole}(m=3,4)  &0.07 & 0.7 \\
m214 &{\tt jaffe}+{\tt gals}+{\tt pert}+{\tt mpole}(m=3,4,5)  &0.04 & 0.4 \\
     \hline
m221 &{\tt jaffe}+{\tt ahern}+{\tt gals}+{\tt pert}   &0.10 & 1.0 \\
m222 &{\tt jaffe}+{\tt ahern}+{\tt gals}+{\tt pert}+{\tt mpole}(m=3)  &0.08 & 0.8\\
m223 &{\tt jaffe}+{\tt ahern}+{\tt gals}+{\tt pert}+{\tt mpole}(m=3,4)  &0.06 & 0.6 \\
m224 &{\tt jaffe}+{\tt ahern}+{\tt gals}+{\tt pert}+{\tt mpole}(m=3,4,5) &0.05 & 0.5 \\
  \hline
  \hline
\end{tabular}}
\begin{tablenotes}
\footnotesize
\item[*] {For Model components, {\tt anfw} refers to the NFW profile for the halo, {\tt jaffe} refers to the pseudo-Jaffe profile for the halo,  {\tt gals} refers to being the BCG treated same as the other member galaxies, {\tt ahern} refers to the Hernquist profile for the BCG, {\tt pert} refers to an external shear, {\tt mpole} refers to multipole perturbations of order m. See the text for a detailed explanation on the label.}
\end{tablenotes}
  %\end{center}
\end{threeparttable}
\end{table}

As the first system of a quasar lensed by a cluster, the large-separation gravitationally lensed quasar
SDSS J1004+4112 \cite{2003Natur.426..810I,2004ApJ...605...78O,2005PASJ...57L...7I}
was discovered from the Sloan Digital Sky Survey (SDSS) \cite{2000AJ....120.1579Y,2004AJ....128..502A} data, which has generated much interest due to its rarity and particularity.
This system is composed of five multiple images of a quasar at $z_{\rm s}=1.734$ \cite{2008PASJ...60L..27I} with the maximum separation angle between the multiple images of 14.62$''$. A cluster of galaxies as the lensing object has been identified at $z_{\rm l}=0.68$ centered among the five images, the brightest galaxy and more galaxy cluster members have also been observed. In Fig.~\ref{Fig1}, we present the five multiple images of the lensed quasar as well as seven lensed background galaxies at three different redshifts, which have also been spectroscopically confirmed \cite{2005ApJ...629L..73S,2009MNRAS.397..341L,2010PASJ...62.1017O}. 
%We also present images of lensed background galaxies at redshifts 3.33, 2.74 and 3.28 in Fig.~\ref{Fig1}.
Such a wealth of observational constraints suggest that it may be possible to tightly constrain the Hubble constant from this lens system.

In our investigation, we employ positions of the five quasar images in Ref.~\cite{2010PASJ...62.1017O} which was adapted according to the Hubble Space Telescope Advanced Camera for Surveys (HST/ACS) F814W images \cite{2005PASJ...57L...7I}. For images of lensed background galaxies, we include the constraints identified in Refs.~\cite{2005ApJ...629L..73S,2009MNRAS.397..341L,2010PASJ...62.1017O}. There are 7 multiply-imaged knots originating from three galaxies (A, B, and C) at different redshifts. The galaxy A contains three knots, each of which is lensed into five images. Galaxies B and C each have two knots in a galaxy with three lensed images. In Table~\ref{table:1}, we summarize positions of the multiple images, which are distributed in a very wide range of radius. This remarkable feature helps us model this cluster strong lensing system in great detail and to constrain the Hubble constant. Moreover, the newly measured time delay between lensed quasar images D and C ($t_{\rm DC}=2456.99\pm5.55$ days) is included in our analysis in addition to the previously measured time delays between images A-C and B-C \cite{2022ApJ...937...34M,2007ApJ...662...62F,2008ApJ...676..761F} as well as magnitude differences between the quasar multiple images.
For the brightest cluster galaxy (BCG), the position is fixed
to the observed position at ($\Delta x$, $\Delta y$)=(7.104$''$, 4.362$''$) with respect to the quasar image A. We also include the positions, ellipticities, position angles and luminosity ratios (with respect to the BCG) of 90 cluster galaxy members, which are selected from the HST/ACS F435W and F814W images with the red-sequence feature and whose properties are measured also in the HST/ACS F814W image using the \textsc{SExtractor} software \cite{1996A&AS..117..393B}.

\section{Lens Modeling} \label{sec:model}

We compute the lens model with the publicly available software \textsc{glafic} \cite{2010PASJ...62.1017O,2021PASP..133g4504O}, which is designed to model strong lens systems and can compute efficiently lensed image positions of point sources with adaptive-meshing.

\subsection{Dark matter halo}

The Navarro-Frenk-White (NFW) profile is employed to model the mass
distribution of a dark matter halo \cite{1997ApJ...490..493N}. The radial density profile of the NFW profile is described by
\begin{equation}
\rho=\frac{\rho_{\rm s}}{(r/r_{\rm s})(1+r/r_{\rm s})^2},
\end{equation}
where $r_{\rm s}$ indicates the scale radius and $\rho_{\rm s}$ is the characteristic density. The projected NFW profile that is used for lensing analysis is parameterized by mass $M$, the position $x$ and $y$, the ellipticity $e$, the position angle $\theta_{\rm e}$, and the concentration parameter $c\equiv r_{\rm vir}/r_{\rm s}$.
In our analysis, we optimize all these parameters with the range for the mass of  $10^{14}-10^{15}h^{-1}M_{\odot}$ and concentration parameter from $1.0$ to $10.0$. We adopt the fast approximation of the lensing  calculation of the NFW profile proposed in Ref.~\cite{2021PASP..133g4504O}, which is named {\tt anfw} in \textsc{glafic}.

In addition to the NFW profile, the Pseudo-Jaffe Ellipsoid ({\tt jaffe} in \textsc{glafic}) profile \cite{1983MNRAS.202..995J,2001astro.ph..2341K} is also used to model the dark matter halo.
In this model, the radial density profile is described as 
\begin{equation}
\rho=\frac{\rho_\mathrm{s}}{\{1+(r/r_{\rm core})^2\}\{1+(r/r_{\rm trun})^2\}}.
\end{equation}
The projected density profile is parameterized  by the  velosity dispersion $\sigma$, the position, the ellipicity, the position angle, the core radius, and the truncation radius.
In our analysis, we optimize all the parameters, but we restrict the range of $\sigma$ from $500\kms$ to $800\kms$,  
 $r_{\rm trun}$ from $5.0^{''}$ to $50.0^{''}$ and $r_{\rm core}$ is smaller than $15.0^{''}$.  

\subsection{The brightest cluster galaxy and other cluster galaxy members}

In some of our mass models, the brightest cluster galaxy (BCG) is parameterized by the Hernquist profile \cite{1990ApJ...356..359H}, which
takes the following radial form
\begin{equation}
\rho=\frac{M_{\rm tot}}{2\pi(r/r_{\rm b})(1+r/r_{\rm b})^3},
\end{equation}
where the scale radius $r_{\rm b}$ is related to the effective radius $R_{\rm e}$ of a project surface mass density as $r_{\rm b}=0.551R_{\rm e}$.
The projected density profile can be parametrized by the total mass $M_{\rm tot}$, the position, the ellipticity, position angle and the scale radius.
In our analysis, we fix the position $x$ and $y$ to the observed BCG centroid as described in Sec.~\ref{sec:data} and we optimize the total
mass $M_{\rm tot}$ and the scale radius $r_{\rm b}$ with the range of
$10^{12}-10^{13}h^{-1}M_{\odot}$ and $1.0^{''}-10.0^{''}$, respectively.
We adopt the fast approximated calculation of lensing properties of the Hernquist profile ({\tt ahern} in \textsc{glafic}) proposed in Ref.~\cite{2021PASP..133g4504O}.
In addition, we add Gaussian priors to the ellipticity and the position angle of the BCG, $e=0.355\pm0.05$ and
$\theta_{\rm e}=-20.47^{\circ}\pm5^{\circ}$, respectively, based on the observed shape of the light profile of the BCG.

For the other cluster galaxy members, the scaled Pseudo-Jaffe ellipsoids ({\tt gals} in \textsc{glafic}) are included to model.
The velocity dispersion $\sigma$ and truncation
radius $r_{\rm trun}$ for each galaxy are scaled based on its luminosity relative to that of the BCG as
\begin{align}
\frac{\sigma}{\sigma_{*}} &=\left(\frac{L}{L_{*}}\right)^{1/4},\\
\frac{r_{\rm trun}}{r_{\rm trun,*}} &=\left(\frac{L}{L_{*}}\right)^{\eta}.
\end{align}
During our optimization,  $\sigma_*$, $r_{\rm trun,*}$, and
$\eta$ are all allowed to vary with the range
for the
$\sigma_*$ of
 $100.0-400.0\kms$,
$r_{\rm trun,*}$ from $1.0^{''}$ to $100.0^{''}$ and $\eta$ from $0.2$ to $1.5$.

To check the dependence of our results on the treatment of the BCG, in some lens mass models we do not explicitly include the BCG as a separate component, but include it as one of Pseudo-Jaffe ellipsoids of cluster member galaxies ({\tt gals}) with its velocity dispersion and the truncation radius determined following the same scaling relations for cluster member galaxies.

\subsection{External perturbations and multipole perturbations}
In order to achieve a better fit, multipole perturbations are sometimes included in mass modeling. In this paper, we consider multipole perturbations of 
order up to 5 at most (e.g., $m=2$, $3$, $4$, and $5$) with the potential
\citep{2003MNRAS.345.1351E,2004PASJ...56..253K,2005MNRAS.364.1459C,2006ApJ...642...22Y}
\begin{equation}
\phi=-\frac{\epsilon}{m}r^2\cos m(\theta-\theta_{\epsilon}-\pi/2).
\end{equation}
For each order, $\epsilon$ and $\theta_{\epsilon}$ are model parameters, which we optimize without any prior ranges. The perturbation with $m=2$ is referred to as an external shear \citep{1997ApJ...482..604K} and named {\tt pert} in \textsc{glafic}, while the terms with $m\geq 3$ are named {\tt mpole}.

\subsection{Model optimization}

In this paper, we consider 16 different lens mass models
 to fit this cluster-lensed quasar system in order to check how the constraint on $H_0$ depends on the different assumption on the lens mass model.
Specifically, we consider the following three variation of assumptions on the mass model, 
(1) which model is used to describe the dark matter halo;
(2) whether to remove the BCG from all cluster member galaxies and include a separate Hernquist
component to model it;
(3) how many orders of perturbations are considered. 

More specifically, we use the NFW profile ({\tt anfw} in \textsc{GLAFIC}) and the Pseudo-Jaffe Ellipsoid profile ({\tt jaffe} in \textsc{GLAFIC}) to describe the mass distribution of dark matter halos.
For convenience, we label modeling of the dark matter halo with  {\tt anfw} or {\tt jaffe} as 1 or 2.
In addition, if we include the brightest cluster galaxy (BCG) as one of the cluster members, we apply the scaled Pseudo-Jaffe Ellipsoids  profile ({\tt gals} in \textsc{GLAFIC}) to model  mass distribution; we label this situation as 1. On the contrary, BCG as a separate Hernquist profile ({\tt ahern} in \textsc{GLAFIC}) component from the other cluster member galaxies is labeled as 2.
As for perturbations, we consider not only external shear ({\tt pert} in \textsc{GLAFIC}) but also multipole perturbations ({\tt mpole} in \textsc{GLAFIC}) of order up to 5 at most. 
Conveniently,
labels 1-4 are used to represent which
corresponds to {\tt pert} only, {\tt pert} plus {\tt mpole} with $m=3$, {\tt pert} plus {\tt mpole} with $m=3$, $4$, and {\tt pert} plus {\tt mpole} with $m=3$, $4$, $5$, respectively.

We summarize all our mass models and
corresponding labels, which describe the difference of the models more concisely, in Table~\ref{table:2}. The label m$ijk$ means the label $i$ for the dark matter halo, the label $j$ for the cluster member galaxies (BCG and other members), and the label $k$ for perturbations. 
For instance, the label `m112' means that we employ the NFW profile ({\tt anfw}) to model the dark matter halo, include BCG as one of the cluster members that are described by the scaled Pseudo-Jaffe Ellipsoid profile ({\tt gals}), and the external shear ({\tt pert}) and third-pole perturbation ({\tt mpole}(m=3)) are included. Therefore we also note it as  model m112 ({\tt anfw}+{\tt gals}+{\tt pert}+{\tt mpole}(m=3)).  
As another example, the difference of the model m122 ({\tt anfw}+{\tt ahern}+{\tt gals}+{\tt pert}+{\tt mpole}(m=3)) with the model m112 is that we
describe the BCG separately with the Hernquist profile ({\tt ahern}) instead of treating the BCG  similarly with other cluster member galaxies by the scaled  Pseudo-Jaffe Ellipsoid profile ({\tt gals}).
As yet another example, the difference of the model m222 ({\tt jaffe}+{\tt ahern}+{\tt gals}+{\tt pert}+{\tt mpole}(m=3)) with the model m122 is that 
the Pseudo-Jaffe Ellipsoid profile ({\tt jaffe}) is used to parameterised the dark matter halo instead of the NFW ({\tt anfw}) profile. If we include an additional order of multipole perturbations to the model m222, the new model is referred to as m223 ({\tt jaffe}+{\tt ahern}+{\tt gals}+{\tt pert}+{\tt mpole}(m=3,4)). 

In cluster strong lens modeling, it is customary to assume positional errors that are much larger than measurement errors of positions of multiple images, in order to take account of the complexity of cluster mass distributions due to e.g., substructures in clusters and the line-of-sight matter fluctuations. For massive clusters, a typical scatter of multiple image positions due to such complexity is $\sim 0.5''-1''$ (e.g., \cite{2018ApJ...863...60R}). Because of the lack of the guidance on what positional errors should be assumed for SDSS J1004+4112, we choose positional errors for each of our mass models such that the value of reduced-$\chi^2$ for the best-fitting model becomes roughly equal to one. Following \cite{2010PASJ...62.1017O}, we also choose smaller positional errors for quasar multiple images than for galaxy multiple images, because accurate quasar image positions are important for accurate predictions of time delays between multiple images. Table~\ref{table:2} summarizes specific positional errors that we adopt for the 16 different lens mass models.

In our mass modeling, we have 71 observational constraints. Naturally, different mass models use different model parameters, and as a result the degree of freedom (dof) is different for different mass models.
For example, for the model m111 ({\tt anfw}+{\tt gals}+{\tt pert}) in which we use the NFW profile to describe the
dark matter halo, do not employ the Hernquist profile  to model the BCG separately from all the cluster member galaxies, and just consider the external shear
(i.e., {\tt pert}), there are 28 model parameters and the degree of freedom is 43.
For model m224 ({\tt jaffe}+{\tt ahern}+{\tt gals}+{\tt pert}+{\tt mpole}(m=3,4,5)) in which 
the dark matter halo is parameterised by the Pseudo-Jaffe Ellipsoid profile, the Hernquist
profile is employed to model the BCG, and we not
only consider external shear but also third-, fourth-, and fifth-order multipole perturbations, we have 39 model parameters leaving with 32 degree of freedom.

In our all calculations and model parameter optimizations, we use a standard $\chi^2$ minimization method implemented in the \textsc{glafic} software and estimate $\chi^2$ in the source plane by taking the full account of the magnification tensor at each image position (see \cite{2010PASJ...62.1017O}) in order to speed up the calculations.

\begin{table}
  \caption{
  The Hubble constant with 68.3$\%$ confidence interval and reduced-$\chi^2$ ($\chi^2$/dof) values for the 16 different lens mass models.  
  \label{table:3}}
  \vspace{3mm}
  %\begin{center}
  \setlength{\tabcolsep}{6.5mm}{
    \begin{tabular}{ccc}
     \hline\hline
Model   & $H_0$ & $\chi^2$/dof \\
& [$\Mpc$] & \\
     \hline
m111 & $71.4^{+5.0}_{-5.0}$ &46.7/43\\
m112 & $59.2^{+1.5}_{-3.5}$ &36.0/41\\
m113 & $57.5^{+2.0}_{-3.0}$ &41.6/39\\
m114 & $58.0^{+2.0}_{-3.0}$ &31.1/37\\
    \hline
m121 & $71.1^{+2.0}_{-3.0}$ &38.2/39\\
m122 & $62.4^{+3.5}_{-3.5}$ &37.5/37\\
m123 & $60.9^{+2.5}_{-2.0}$ &38.7/35\\
m124 & $60.8^{+2.0}_{-2.0}$ &31.0/33\\
     \hline
m211 & $88.9^{+4.0}_{-3.0}$  &53.8/42\\
m212 & $88.7^{+3.5}_{-3.0}$ &39.1/40\\
m213 & $79.5^{+4.0}_{-2.5}$ &38.6/38\\
m214 & $79.6^{+3.0}_{-3.0}$ &38.1/36\\
     \hline
m221 & $76.7^{+3.5}_{-3.5}$ &45.8/38\\
m222 & $74.8^{+3.5}_{-3.5}$ &42.8/36\\
m223 & $64.2^{+3.5}_{-3.5}$ &39.6/34\\
m224 & $62.1^{+3.0}_{-2.5}$  &29.7/32\\
%     \hline
%combined all lens models &  65.10 & $^{+17.15}_{-6.61}$\\
%combined best fitted lens models & 59.35 & $^{+3.36}_{-3.47}$\\
  \hline
  \hline
\end{tabular}}
  %\end{center}
\end{table}

\begin{figure}
\begin{center}
\includegraphics[width=1.0\linewidth]{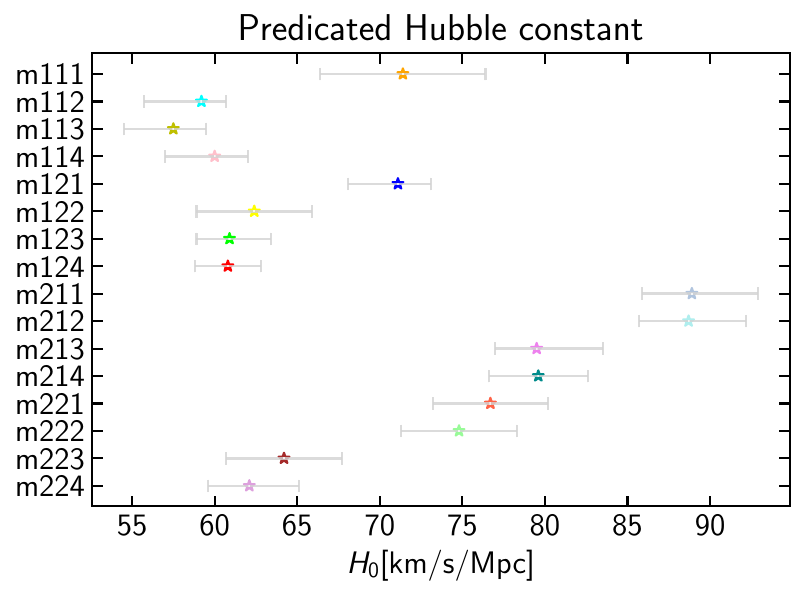}
\includegraphics[width=1.0\linewidth]{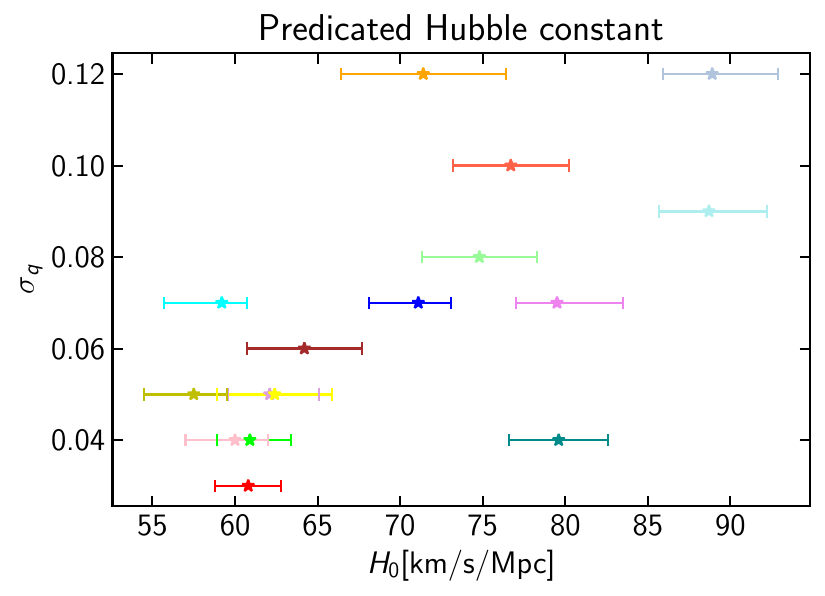}
\end{center}
\caption{ {\it Top:} The summary of the constraint on the Hubble constant for all the 16 different lens mass models. 
%The combined result with equal
%weighting of all these measurements is also shown with black star and grey bar.
{\it Bottom:} Similar to the top panel, but for assumed position errors of quasar images $\sigma_{\rm q}$ in 16 different lens mass
models with measurements of the Hubble constant.
}
\label{Fig2}
\end{figure}

\begin{figure}
%\centering
\includegraphics[width=0.97\linewidth]{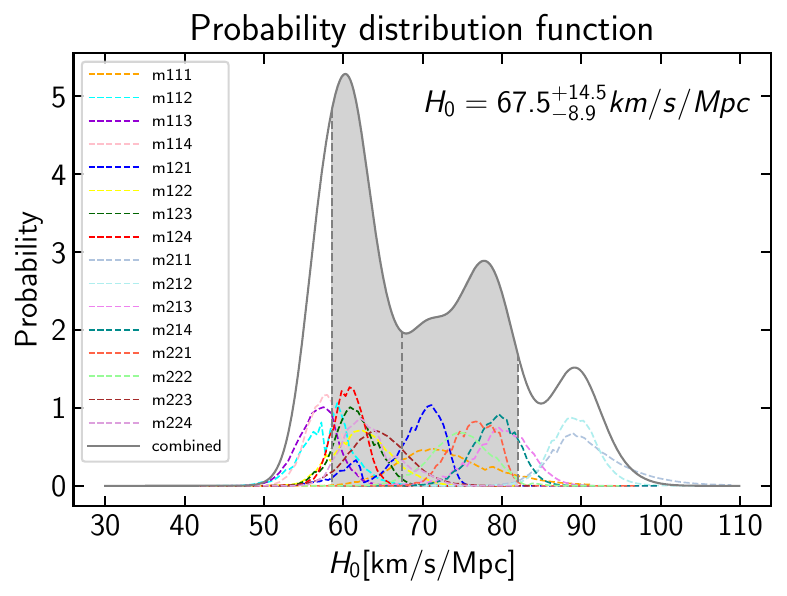}
\includegraphics[width=0.97\linewidth]{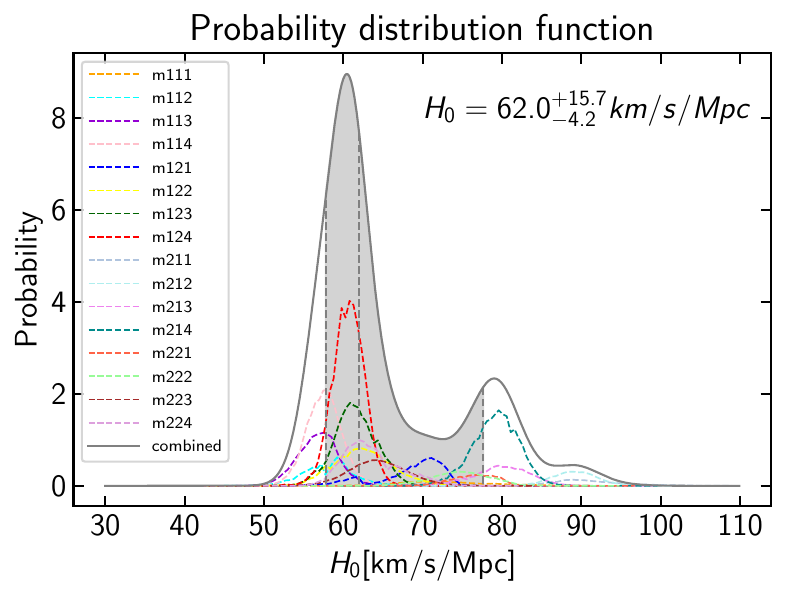}
\includegraphics[width=0.97\linewidth]{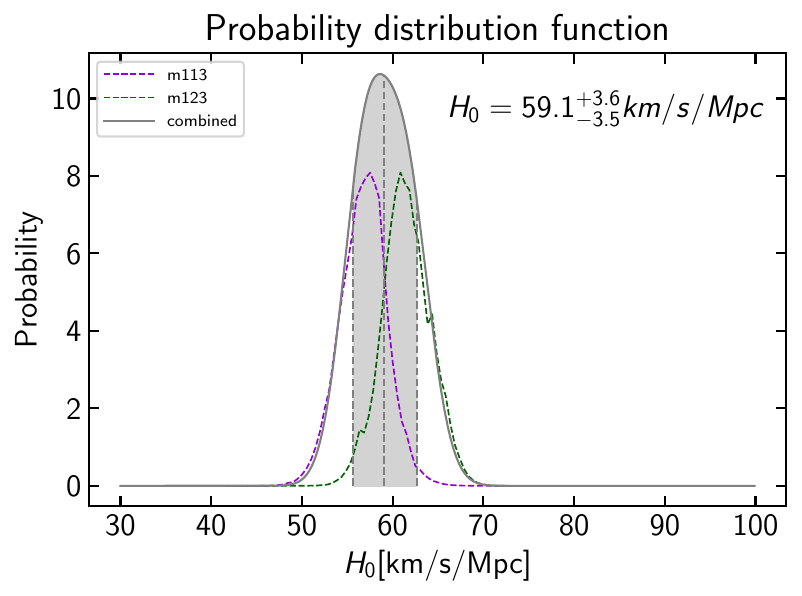}
\caption{{\it Top:} The probability distribution functions (PDFs)
of the Hubble constant
for all the 16 different lens mass models ({\it dashed}), as well as the PDF after combining all the 16 mass models with equal weighting ({\it solid}). The vertical gray dotted lines and shaded region indicate the median and $68.3\%$ confidence interval for the PDF of the combined result.
{\it Middle:} Similar to the top panel, but combining all the mass models with different weighting.
{\it Bottom:} Also similar to the top panel, but for 2 best mass models that best reproduce the observed shapes of the lensed quasar host galaxy and the lensed galaxies behind the cluster. }
\label{Fig3}
\end{figure}

\section{Hubble constant from the time delay cosmology}\label{sec:result}

\begin{figure*}
%\centering
\includegraphics[width=8.4cm]{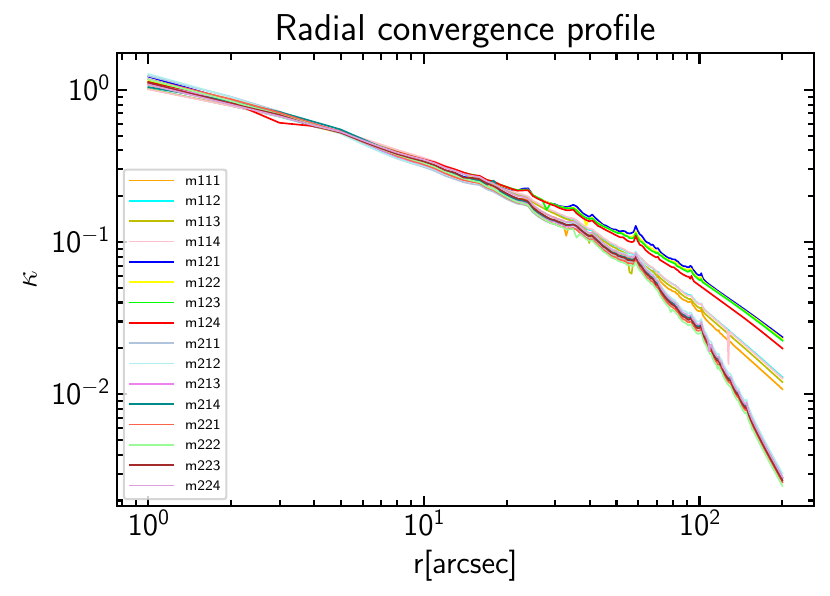}
\includegraphics[width=8.4cm]{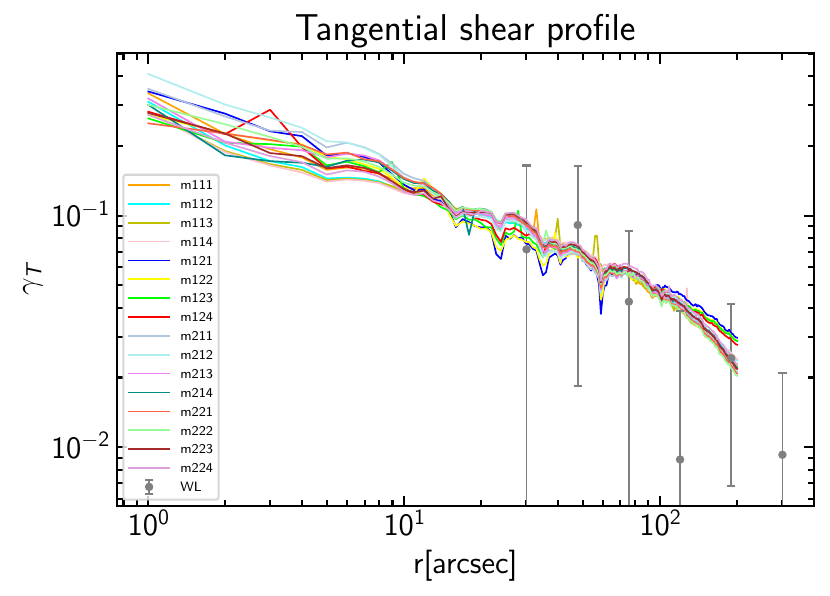}
\caption{Radial convergence ({\it left}) and tangential
shear ({\it right}) profilers for the 16 different lens mass models, which are computed assuming the source redshift of $z_{\rm s}=1$.
Grey dots with errors represent tangential shear measurements from previous weak gravitational leasing analysis \cite{2012MNRAS.420.3213O}.}
\label{Fig4}
\end{figure*}

Thanks to the progress of observations as well as theoretical understanding of the lens model dependence, 
time-delay cosmography has emerged as an important independent technique to constrain the Hubble constant. 
For a strong lensing system with a background
quasar as a source and a cluster as a lens,
variability patterns of a quasar for different multiple images have some time lag because of different light
paths for the different multiple images.
The time delay between image $i$ and $j$ is calculated as (e.g., \cite{1985A&A...143..413S})
\begin{equation}
\Delta t_{ij}=\frac{D_{\Delta{\rm t}}}{c}\Delta\phi_{ij},
\end{equation}
where the time delay distance $D_{\Delta{\rm t}}$ is defined as (e.g., \citep{1964MNRAS.128..307R,1992grle.book.....S})
\begin{equation}
D_{\Delta{\rm t}}\equiv(1+z_{\rm l})\frac{D_{\rm l}^{\rm A}D_{\rm s}^{\rm A}}{D_{\rm ls}^{\rm A}},
\end{equation}
with 
$z_{\rm l}$ being the redshift of the lens,
$D_{\rm l}^{\rm A}$, $D_{\rm s}^{\rm A}$, and $D_{\rm ls}^{\rm A}$ being 
angular diameter distances from the observer to the
lens, from the observer to the source, and between the lens and the source, respectively.
Representing the angular image position in the image plane and the angular source position in the source plane by $\vec{\theta}$ and $\vec{\beta}$, respectively, the Fermat potential difference
$\Delta\phi_{ij}$ is calculated as (e.g., \cite{1986ApJ...310..568B})
\begin{equation}
\Delta\phi_{ij}=\phi(\vec{\theta}_i;\vec{\beta})-\phi(\vec{\theta}_j;\vec{\beta}),
\end{equation}
\begin{equation}
\phi(\vec{\theta};\vec{\beta})=\frac{1}{2}(\vec{\theta}-\vec{\beta})^2-\psi(\vec{\theta}),
\end{equation}
where $\psi(\vec{\theta})$ indicates the gravitational lens potential.

Given the definition of the angular diameter distance
in the Friedmann-Lemaitre-Robertson-Walker (FLRW) metric
\begin{equation}
D^{\rm A}(z_{\rm 1},z_2)=\frac{1}{1+z_2}f_{\rm K}[X(z_1,z_2)],
\end{equation}
where the spatial curvature is described by $K$,
\begin{equation}\label{eq12}
f_{\rm K}(X) = \left\lbrace \begin{array}{lll}
K^{-1/2}\sin(K^{1/2}X),~~~~~~~~~~~~K>0\\
X,~~~~~~~~~~~~~~~~~~~~~~~~~~~~~~~~~K=0 \\
(-K)^{-1/2}\sinh[(-K)^{1/2}X],~K<0,\\
\end{array} \right.
\end{equation}
and 
\begin{equation}
X(z_1,z_2)=\frac{c}{H_0}\int_{z_1}^{z_2} E(z')dz',
\end{equation}
with the dimensionless Friedman equation being $E(z)$,
the Hubble constant is associated with time delays
in the flat Universe ($K=0$) as follows
\begin{equation}
H_0=\frac{1}{\Delta t_{ij}}\frac{\int_{0}^{z_{\rm l}} E(z')dz'\int_{0}^{z_{\rm s}} E(z')dz'}{\int_{z_{\rm l}}^{z_{\rm s}} E(z')dz'}\Delta\phi_{ij}.
\end{equation}
%\mo{It looks to me (1+zl)/(1+zs) is not needed, so I removed it. Please double check.}
It is clearly seen that $H_0$ can be constrained from the measurements of time delays, as long as the Fermat potential $\Delta\phi_{ij}$ is well constrained from lens mass modeling. 
Throughout our analysis, we assume a flat Universe with $\Omega_{\rm m}=0.3$ and $\Omega_{\Lambda}=0.7.$

In practical analysis, we include $H_0$ as a model parameter and vary it together with all the other model parameters to simultaneously fit the image positions, magnitude differences, and time delays that are summarized in Table~\ref{table:1}. The error on $H_0$ is estimated by changing the $H_0$ value around the best-fit value with a step size of $0.5\Mpc$, and for each value of $H_0$ we optimize the other model parameters to obtain the best-fit $\chi^2$, and derive the $1\sigma$ error by the range where the difference of the best-fit $\chi^2$, $\Delta\chi^2$, is smaller than $1$, assuming Gaussian errors.

We summarize our results in Table~\ref{table:3} and Fig.~\ref{Fig2}. We find that the best-fit $H_0$ values are significantly different for different lens mass models and are much larger than the statistical errors derived for individual lens mass models. For instance, when the NFW profile is used to model the dark matter halo and using the external shear to describe the perturbation (m111), we obtain $H_0=71.4^{+5.0}_{-5.0}\Mpc$. When more multipole perturbations are added, it becomes $H_0=59.2^{+1.5}_{-3.5}\Mpc$ for m112, and $H_0=57.5^{+2.0}_{-3.0}\Mpc$ for m113. We find that the best-fit Hubble constant becomes smaller when more multipole perturbations are added. Interestingly, this trend appears to be common for all the mass models we consider in this paper, and is also seen in the bottom panel of Fig.~\ref{Fig2}. 

We find that changing the halo density profile from the NFW profile to the Pseudo-Jaffe Ellipsoid profile has a larger impact on the resulting $H_0$ value. For instance, we obtain $H_0=79.6^{+3.0}_{-3.0}\Mpc$ in lens modeling
combination m214, which is significantly larger than $H_0$ for m114. Fig.~\ref{Fig2} clearly indicates that $H_0$ values tend to be much higher in general when the Pseudo-Jaffe Ellipsoid profile is used to model the dark matter halo. Modelling the BCG separately as the Hernquist profile also has some impact on the result, e.g., $H_0=62.1^{+3.0}_{-2.5}\Mpc$ is derived from m224, which is  much smaller than the $H_0$ value obtained for m214.

Combining results from these different lens mass models require the determination of how the weight is assigned to each lens mass model. For instance, in the analysis of SN Refsdal \cite{2023Sci...380.1322K}, a weight for each mass model is assigned based on the reproducibility of flux ratios and time delay ratios of multiple supernova images that are not included in the mass modeling. In our case, we cannot use the flux ratios and time delay ratios for weighting as they are explicitly included as model constraints. Given the lack of the clear guidance on how to assign weights to individual lens mass models, first we simply sum up posteriors $\exp(-\Delta\chi^2/2)$ of the Hubble constant from individual mass models with equal weighting. 
The combined posterior probability distribution function (PDF) of $H_0$, as well as PDFs for individual lens mass models, are shown in Fig.~\ref{Fig3}. The constraint on $H_0$ from this combined PDF is $H_0=67.5^{+14.5}_{-8.9}\Mpc$, which represents a weak constraint on $H_0$ due to the large scatter among different lens mass models. 

To see how the result depends on weighting, we next combine all the lens mass models with weighting based on the assumed positional errors of quasars images $\sigma_{\rm q}$. Specifically, we adopt the weighting
value of  $w_{ i}=1/{\sigma^{2}_{\rm q}}_{,i}$, where ${\sigma_{\rm q}}_{,i}$ are positional errors of individual mass models quoted in Table~\ref{table:2}. With this weighting, we obtain $H_0=62.0^{+15.7}_{-4.2}\Mpc$ (see Fig.~\ref{Fig3}), which represents a slightly tighter constraint than for equal weighting. While the slight improvement comes from the trend of $H_0$ with the positional errors as shown in Fig.~\ref{Fig2}, we conclude that the error on the Hubble constant is still large even with weighting based on positional errors of quasar images.

We note that the 16 lens mass models explored in this paper do not cover the full lens model uncertainty, and combining these 16 lens mass model results with equal weighting or weighting with positional errors is also not well motivated, indicating that this constraint on $H_0$ should be taken with caution. The multiple peaks in the PDF indeed imply that we do not yet fully marginalize over the full lens model uncertainty. Nevertheless, our analysis at least suggests that the lens mass model uncertainty is quite large for the current observational constraints and the treatment of positional errors. 
These findings motivate us to further discuss other possibilities to distinguish or exclude some of the lens mass models to further improve constraints on the Hubble constant.

\begin{figure*}
\centering
\includegraphics[width=4.2cm]{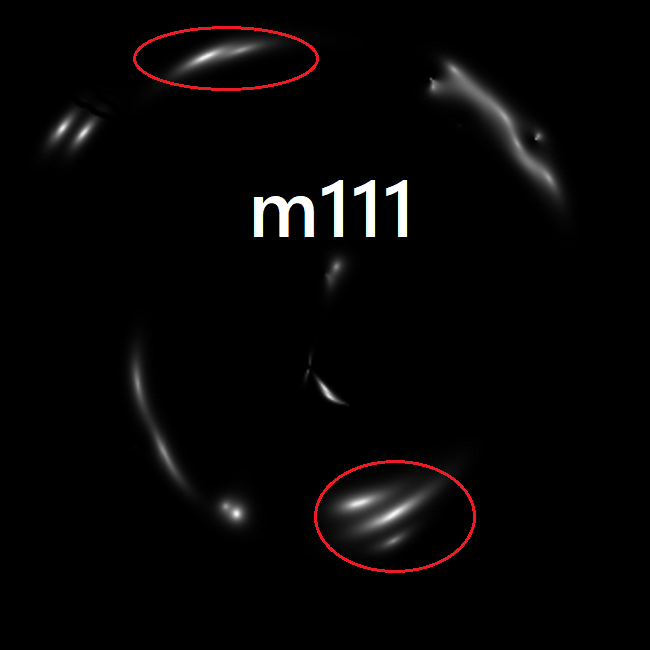}
\includegraphics[width=4.2cm]{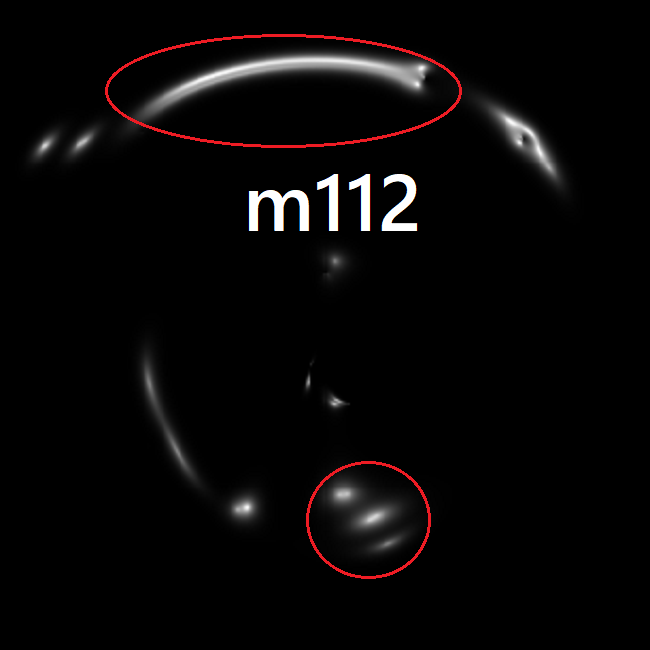}
\includegraphics[width=4.2cm]{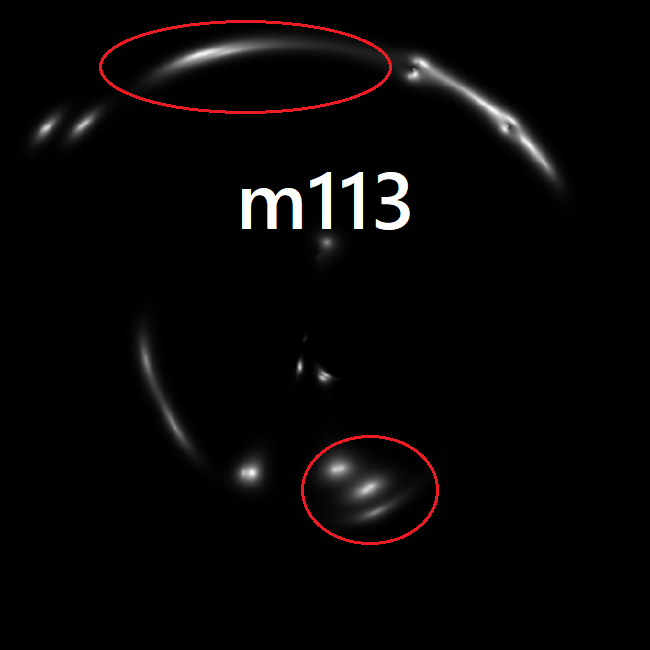}
\includegraphics[width=4.2cm]{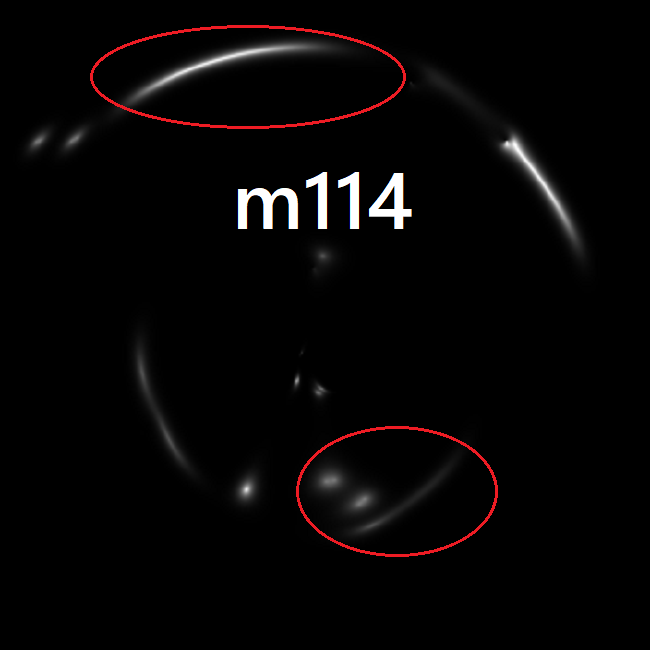}
\includegraphics[width=4.2cm]{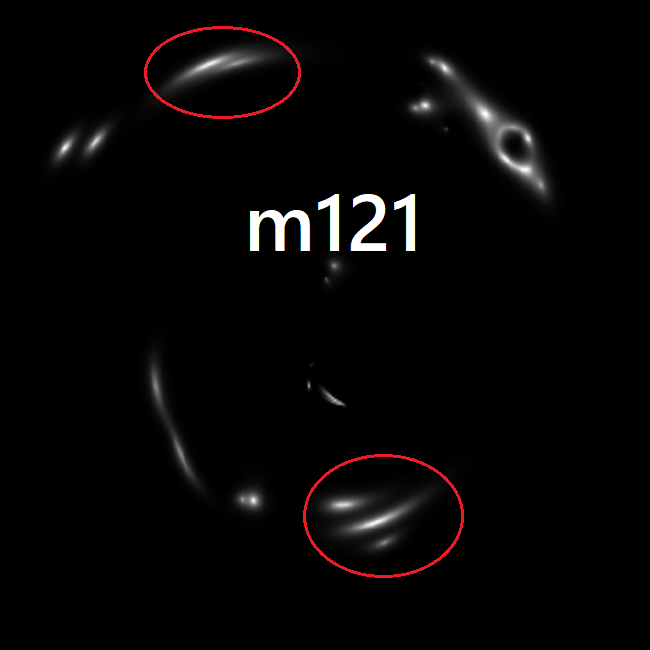}
\includegraphics[width=4.2cm]{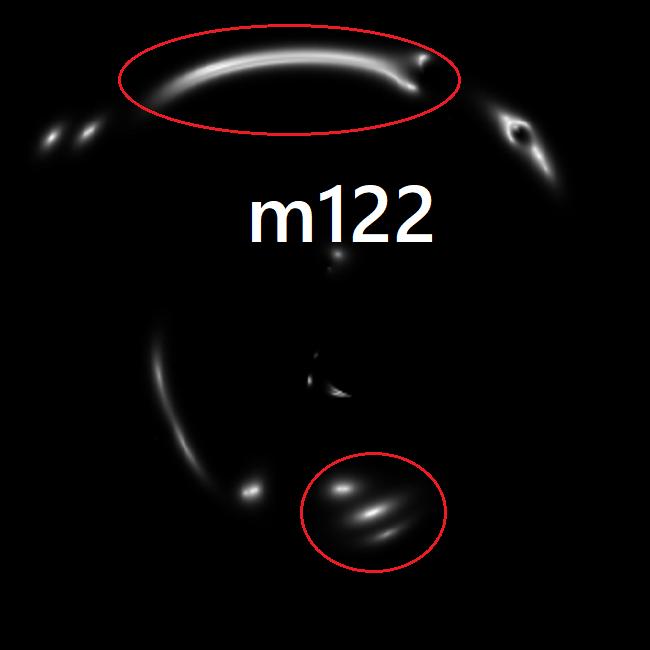}
\includegraphics[width=4.2cm]{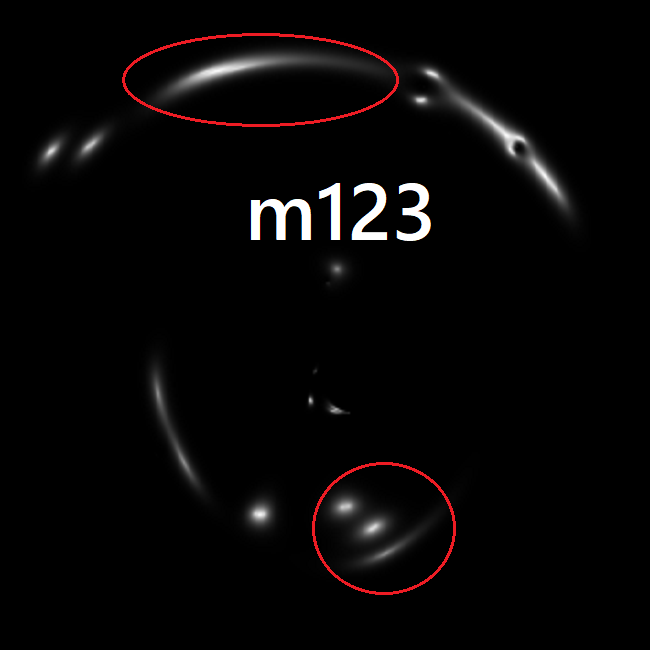}
\includegraphics[width=4.2cm]{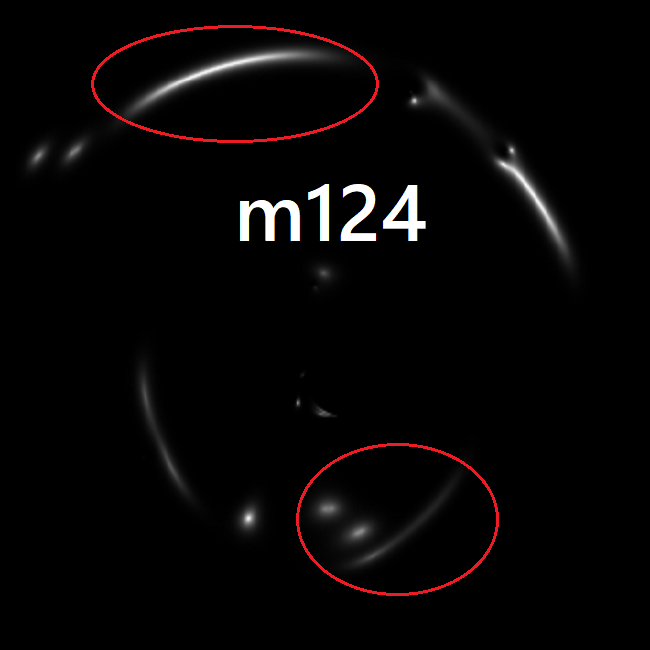}
\includegraphics[width=4.2cm]{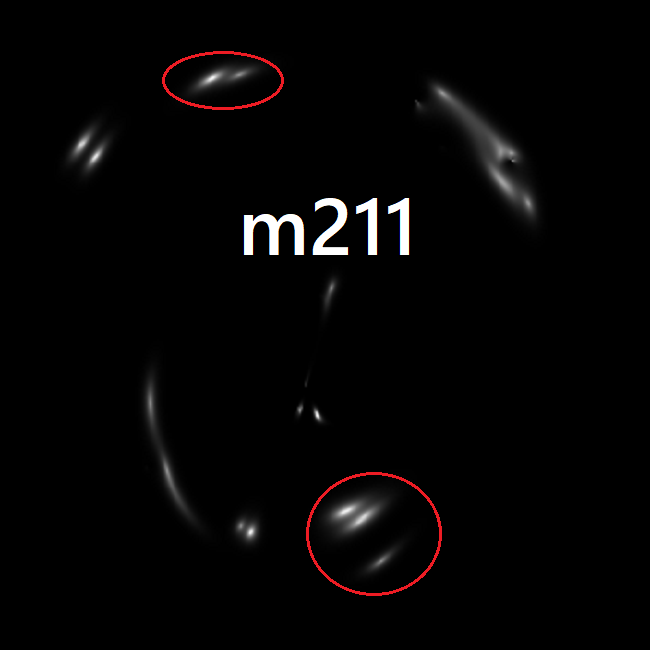}
\includegraphics[width=4.2cm]{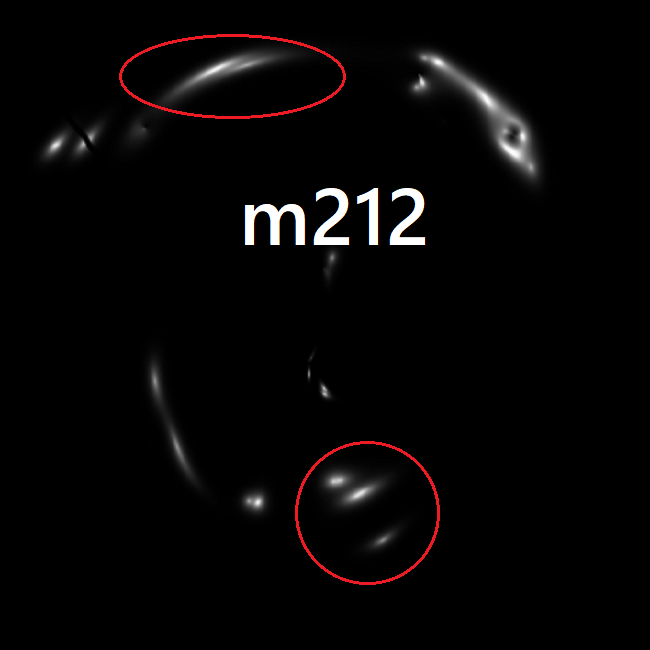}
\includegraphics[width=4.2cm]{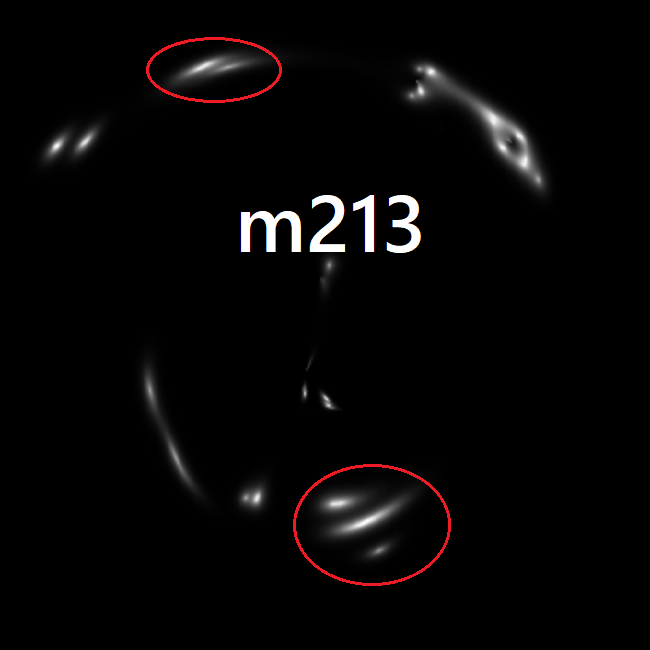}
\includegraphics[width=4.2cm]{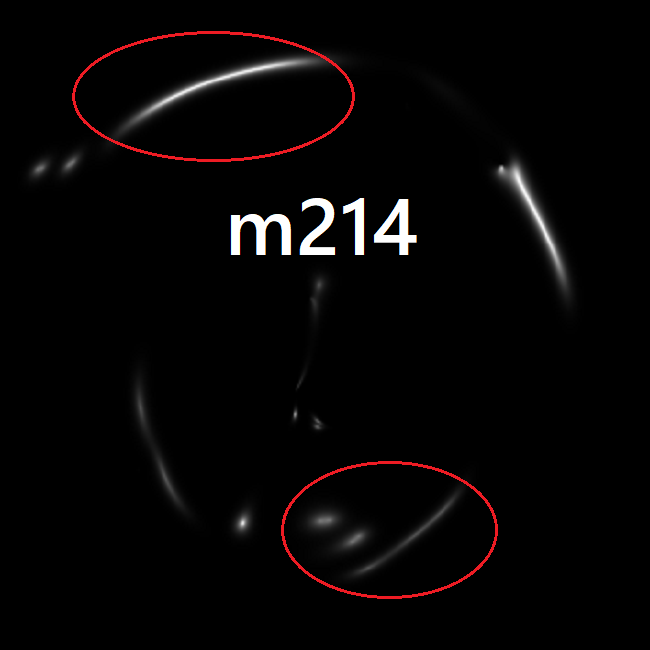}
\includegraphics[width=4.2cm]{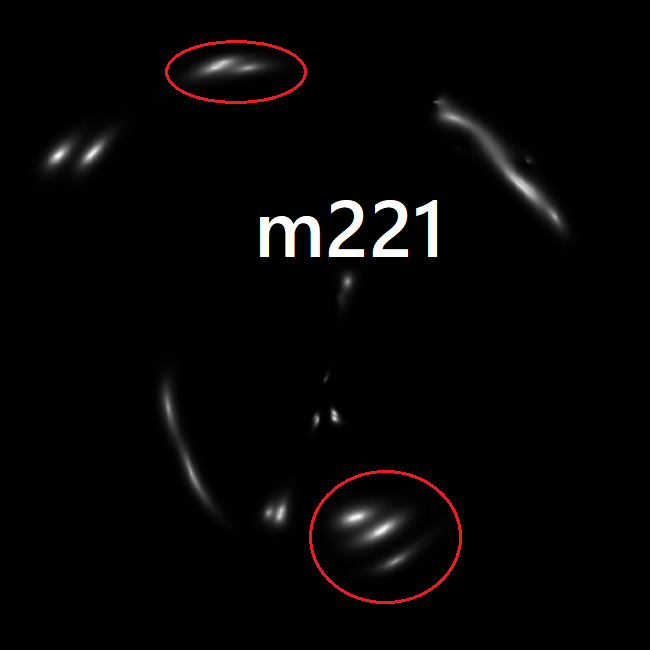}
\includegraphics[width=4.2cm]{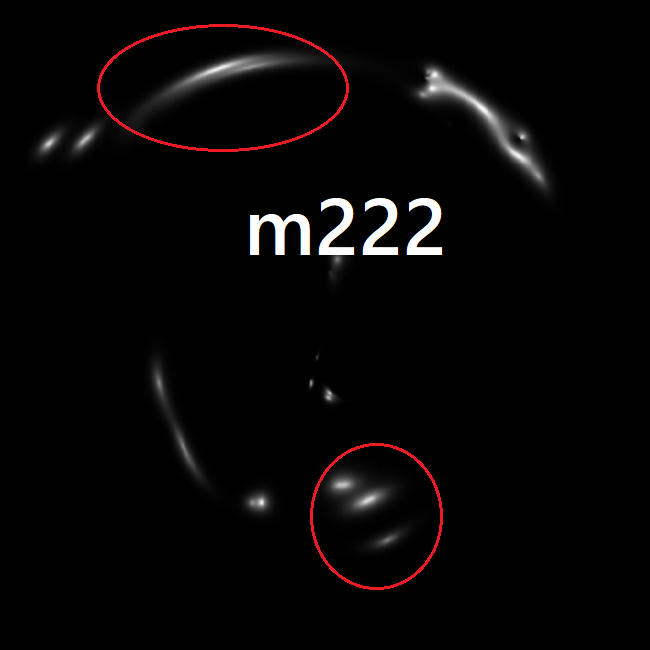}
\includegraphics[width=4.2cm]{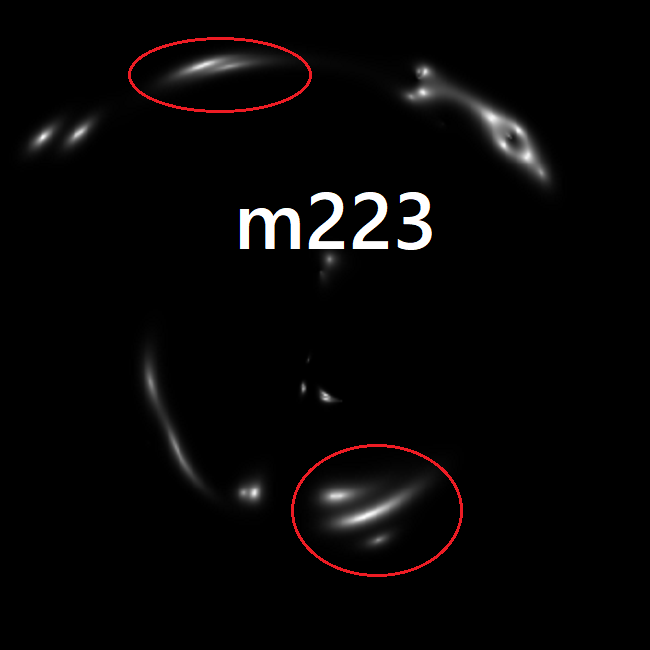}
\includegraphics[width=4.2cm]{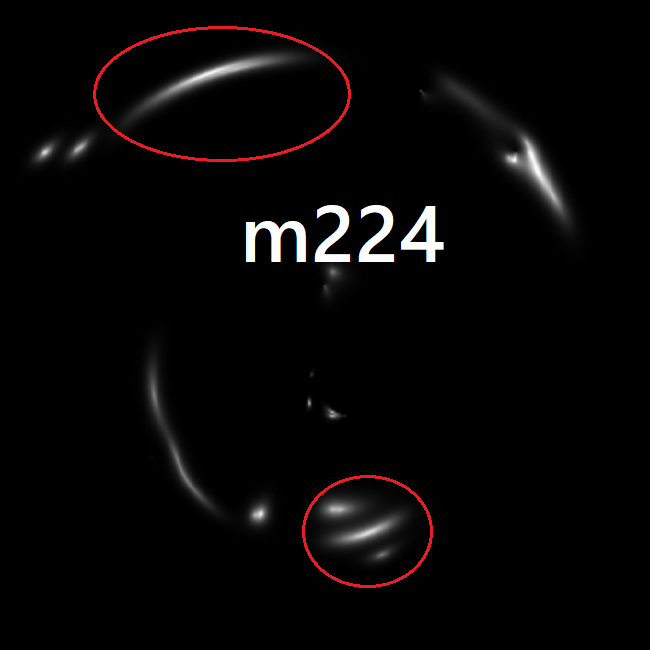}
\caption{The predicated shapes of the lensed host galaxy of the quasar and lensed background galaxies for all the 16 different lens mass models. Shapes enclosed in red ellipses indicate arcs whose shapes are relatively more sensitive to the difference of lens mass models.}
\label{Fig5}
\end{figure*}

\section{Possibility of improving Hubble constant with additional constraints}\label{sec:discussion}

First, it is possible that different lens mass models predict quite different outer density profiles of the lensing cluster, which can be constrained by weak gravitational lensing. To explore this possibility, we calculate radial convergence and tangential shear profiles of all the 16 different lens mass models. 
Since the mass distribution is not circular symmetric, we compute the azimuthal average of the convergence at each radius to derive radial convergence and tangential shear profiles. 
The results shown in Fig.~\ref{Fig4} indicate that the convergence and tangential shear profiles are quite similar between different lens mass models, implying that they cannot be distinguished by weak gravitational lensing. For comparison, we plot results from the weak gravitational lensing analysis in Ref.~\cite{2012MNRAS.420.3213O} using Subaru Suprime-Cam images, which indicates that measurements errors are much larger compared with the difference of tangential shear profiles among different models. 
In order to distinguish these different models with weak lensing, we need much improved weak lensing measurements with deep imaging from space telescopes such as James Webb Space Telescope, and hence is challenging.

In addition, we predict the shapes of lensed quasar host galaxies and lensed background galaxies, which are shown in Fig.~\ref{Fig5}.
We find that there are obvious differences of those shapes between different lens mass models. For instance, as multipole perturbations increase, the arcs become more  extended in most cases. There are also notable differences between different models to describe the dark matter halo such that the arcs are smaller when using the Pseudo-Jaffe Ellipsoid profile than using the NFW profile. In addition, 
some slight differences are presented when we model the BCG with the Hernquist profile separately.
By comparing our models prediations with observations, we find that the lens mass models m113 and m123 best match the observed shapes of lensed quasar host galaxy and lensed galaxies.
%Since the observed arc is not that elongated (see Fig.~\ref{Fig1}), this  result may imply that the NFW model is more accurate in describing the dark matter halo. 

When we restrict our mass models to m113 and m123 and exclude the other models, the combined constraint becomes much tighter, $H_0=59.1^{+3.6}_{-3.5}\Mpc$.
The PDF of $H_0$ for this case is also shown in Fig.~\ref{Fig3}. Again we caution that these two mass models perhaps do not fully cover the lens model uncertainty. However we can at least conclude that adding constraints from shapes of lensed galaxies can greatly help breaking degeneracies between different lens mass models.

\section{Discussions and Conclusion}\label{sec:summary}

In this paper, we explore the possibility of constraining 
the Hubble constant using time-delay cosmography with a cluster-lensed quasar system. As a specific example, 
we focus on the first discovered cluster-lensed quasar lens system SDSS J1004+4112, in which a background quasar is lensed into five multiple images. In addition, several multiple images of galaxies behind the lensing cluster have also been identified. 

In order to check how the constraint on the Hubble constant depends on the choice of the lens mass model,  we employ 16 different lens mass models in which different assumptions on profiles of the dark matter halo, the BCG, and multipole perturbations are made.
We find that the variation of the best-fit values of the Hubble constant is very large among the 16 lens mass models. Interestingly, we find that the value of the best-fit Hubble constant decreases as the complexity of the multipole perturbation increases, whose origin is to be understood.  By summing posteriors of the Hubble constant with equal weighting, the combined constraint is
$H_0=67.5^{+14.5}_{-8.9}\Mpc$.
We also consider different weighting values based on positional errors of quasar images, $w_{ i}=1/{\sigma^{2}_{\rm q}}_{,i}$, to obtain a slightly tighter constraint on $H_0=62.0^{+15.7}_{-4.2}\Mpc$, which does not represent a significant improvement.

We discuss possible ways to distinguish different lens mass models by adding other observational constraints. We find that there are no obvious differences in radial density profiles and tangential shear profiles in the outer region that can be constrained by weak gravitational lensing. In contrast, we find that there are distinct differences in critical curves, and also predicted shapes of quasar host galaxies as well as other lensed galaxies. By comparing those shapes with observations, we find that there are two models (m113 and m123) that best match the observed shapes. By combining the constraints from these two models, we obtain much tighter constraint of $H_0=59.1^{+3.6}_{-3.5}\Mpc$. We emphasize that our analysis with 16 mass models does not fully cover the whole mass model uncertainty and our results are also dependent on how to assign weight to each mass model. Therefore our result should be taken with caution. We expect that our results at least indicate that incorporating as many lensing constraints as possible, most notably those from shapes of lensed extended objects, is important for obtaining tight constraints on the Hubble constant from cluster-lensed quasar lens systems. 

The importance of including constraints from extended arcs has also been emphasized for galaxy-scale quasar lens systems \cite{2001ApJ...547...50K,2010ApJ...711..201S,2017MNRAS.468.2590S}. Our analysis indicates that this may also be the case for the cluster-scale lens system. However, one technical difficulty in the analysis of cluster-scale lens systems lies in the fact that we cannot perfectly reproduced observed image positions in cluster lens modeling, which mainly originates from small-scale matter distributions that cause random scatter in multiple image positions. In presence of significant mismatches of image positions between model predictions and observations, it is not obvious how to rigorously incorporating such shape information in lens modeling, because we cannot simply conduct pixel-by-pixel fitting of lensed arc shapes. How to efficiently and rigorously include such extended arc information in cluster strong lens modeling, which requires understanding of how such small-scale matter distributions not only affect their positions but also their shapes, is an open question, which needs to be overcome for turning cluster-lensed quasar systems into useful and accurate probe of the Hubble constant. In order to obtain accurate constraints on the Hubble constant from time delays, accurate and realistic error bars on time delays are also important.

A caveat is that the situation could be different for more massive clusters (e.g., MACS J1149+2223) where there are many more multiple images, typically more than 100, to tightly constrain lens mass distributions. The large mass model dependence for SDSS J1004+4112 may originate from the small number of multiple images that we can use. Deep imaging of cluster-lensed quasar systems with e.g., James Webb Space Telescope may increase the number of multiple images for SDSS J1004+4112 and other cluster-lensed quasar systems, which can help reduce the lens mass model dependence on the Hubble constant measurement.

\begin{acknowledgments}
%We thank anonymous referees for useful comments and suggestions.
We would like to thank the referee
for useful comments and suggestion.
This work was supported by the National Natural Science Foundation of China under Grants Nos. 12021003, 11690023, and 11920101003; the Strategic Priority Research Program of the Chinese Academy of Sciences, Grant No. XDB23000000; the Interdiscipline Research Funds of Beijing Normal University. YL was supported by the Interdiscipline Research Funds of Beijing Normal University (Grant No. BNUXKJC2017) and China Scholarship Council (Grant No. 202106040084). MO was supported by JSPS KAKENHI Grant Numbers JP22H01260, JP20H05856, and JP22K21349.

\end{acknowledgments}

%\appendix

%\nocite{*}

\bibliography{refer}% Produces the bibliography via BibTeX.

\end{document}